\documentclass{PoS}
\let\ifpdf\relax
\usepackage{verbatim}

\usepackage{bm}
\usepackage{graphicx}
\usepackage{amsmath}
\usepackage{amsthm}
\usepackage{amsfonts}
\usepackage{amssymb}
\newcommand{\tht}{\textheight}
\newcommand{\ig}{\includegraphics}

\mathchardef\minussign="002D

\title{Applications of the Stochastic LapH Method}

\ShortTitle{Appl. of stochastic LapH}

\author{J.~Bulava\\
        CERN, Geneva 23, CH-1211, Switzerland.\footnote{Present address: School of Mathematics, Trinity College, Dublin 2, Ireland. E-mail:\tt{bulavaj@tcd.ie}}\\
        E-mail:\ \email{john.bulava@cern.ch}
}
\author{J.~Foley\\
        University of Utah, Salt Lake City, UT 84112, U.S.A.\\
        E-mail: \email{Justin.Foley@utah.edu}}
\author{\speaker{K.~J.~Juge}\\
        University of the Pacific, Stockton, CA 95211, U.S.A.\\
        E-mail: \email{kjuge@pacific.edu}}
\author{C.~J.~Morningstar\\
        Carnegie Mellon University, Pittsburgh, PA 15213, U.S.A.\\
        E-mail: \email{colin\_morningstar@cmu.edu}}
\author{B.~Fahy\\
        Carnegie Mellon University, Pittsburgh, PA 15213, U.S.A.\\
        E-mail: \email{bfahy@andrew.cmu.edu}}
\author{Y-C. Jhang\\
        Carnegie Mellon University, Pittsburgh, PA 15213, U.S.A.\\
        E-mail: \email{yjhang@andrew.cmu.edu}}
\author{D. Lenkner\\
        Carnegie Mellon University, Pittsburgh, PA 15213, U.S.A.\\
        E-mail: \email{dlenkner@andrew.cmu.edu}}
\author{\speaker{C.~H.~Wong}\\
        University of California, San Diego, CA 92093, U.S.A.\\
        E-mail: \email{rickywong@physics.ucsd.edu}}

\abstract{Progress in computing the hadron spectrum in lattice QCD using stochastic LapH quark propagators is described. The stochastic LapH algorithm is a particular quark smearing algorithm that also allows the computation of all-to-all quark propagators. All-to-all quark propagators are required in our approach of using a large set of spatially extended hadron operators and explicit multi-particle operators to access excited states. We report on the progress made in the various isospin channels on 2+1 dynamical, anisotropic lattices generated by the Hadron Spectrum Collaboration. 
\flushright{CERN-PH-TH/2013-010}

}

\FullConference{The XXX International Symposium on Lattice Field Theory - Lattice2012\\
June 24 - 29, 2012\\
Cairns, Australia}

\begin{document}

\section{Introduction}

One of the goals of our collaboration has been to compute, from first principles lattice QCD simulations, the low-lying hadron spectrum in various symmetry channels in hopes to aid our understanding of QCD through comparisons with low energy effective theories. The reliable extraction of the energies of the excited states in each symmetry channel requires that we use interpolating operators which couple well with the ground states in the channel and also reasonably well with the excited states up to the energy scale that one is interested in. The progress made in the last couple of years for single-particle and some two-particle operators/correlation functions have appeared in various papers and proceedings (\cite{Foley:2012wb}-\cite{Foley:2010vv}). 

The extraction of excited energy levels on dynamical configurations must confront the problem of mixing of single-particle states with multi-particle states and requires new techniques to be developed given the finite resources available. The construction of a set of good interpolating operators which have good signal-to-noise ratio is crucial in computing the excited states through the variational method as the signal for the heavier excited states decay rapidly into the noise, destabilizing the diagonalization procedure. In this contribution, we report on our progress towards solving these issues. 

\section{Lattices} 
The dynamical lattices that we use are the clover-improved, anisotropic lattices (\cite{Edwards:2008ja}) that have been generated by the HadSpec collaboration \cite{Lin:2008pr}. The anisotropy not only allows one to extract energies at early timeslices (as long as the interpolating operators are effective) before the signal-to-noise ratio deteriorates but also generally has smaller fluctuations. There has only been a single lattice spacing generated so far corresponding to $r_0/a_s=3.221(25)$. The anisotropy was tuned to $3.5$ with a measured valued of $3.3$ using the pion dispersion relation. The quark mass used in this study corresponds to pion masses of $240$~MeV (250 configurations). 

\section{Interpolating Operators}
The interpolating operators that are used to form the correlation matrix for digaonalization is constructed so that they transform irreducibly under the lattice symmetry group of rotations and translations and have the appropriate flavor structure and G-parity. Operators for hadrons at rest can be constructed from a handful of conventional quark propagators as long as the displacement types are restricted to the simple displacements. The interpolating operators that we will be using are based on the irreducible representations of the appropriate cubic point groups and the pruning of these types of operators have been studied in detail in previous studies. The choice of the elemental operators to use were based on its simplicity and diversity while maintaining the best overlap and low noise-levels in the single particle correlation functions. A preliminary study of the isovector meson spectrum was presented last year at the lattice conference for the heavier quark mass corresponding to $m_\pi=390\ {\rm MeV}$. 

The same approach is taken in designing/pruning the interpolating operators for hadrons with non-zero momenta, ${\mathbf k}$ (in units of $2\pi/L$). In this study, we consider those states which have momenta of the form, ${\mathbf k}=(0,0,n)$, ${\mathbf k}=(0,n,n)$ and ${\mathbf k}=(n,n,n)$. The group theory for these choices of momenta was discussed in the conference report from last year \cite{Foley:2012wb} and the general discussion of the group theory is given in \cite{Moore:2005dw}. 

\subsection{Quark Propagators}
In order to incorporate hadron operators with non-zero momentum and explicit two-particle operators in our operator set, we construct the hadron interpolating operators from stochastic LapH quark propagators (\cite{Morningstar:2011ka}). The stochastic LapH algorithm is an efficient all-to-all extension of the distillation algorithm (\cite{Peardon:2009gh}) which does not suffer from poor volume scaling behaviour. The method allows one to inject any momentum into the interpolating operators after the propagators have been generated, the construction of extended operators and it also separates the source operator from the sink operator which greatly simplifies the construction of large correlation matrices needed to extract the excited state spectrum. 

\subsubsection{Stochastic LapH Quark Propagators}
In this section, we briefly review the stochastic LapH algorithm. The source vector (for quark label $A$) is given by
$$\varrho_{b\beta k\bar{p}}^{(A)[d]}({\mathbf x},t)=\left(\tilde{D}_k^{(\bar{p})}{\mathbf V}_S{\mathbf P}^{[d]}\varrho^{(A)}\left({\mathbf x},t\right)\right)_{b\beta}$$
where $b$ is the color index, $\beta$ is the spin index, $\tilde{D}_k^{(\bar{p})}$ is the covariant derivative in the $k$ direction with separation $\bar{p}$, ${\mathbf V}_S$ is the matrix of eigenvectors of the Laplacian and ${\mathbf P}^{[d]}$ is the projection operator for the dilution scheme being applied. The dilution schemes can be categorized into four different types:
\[ \begin{array}{lll}
{\mathbf P}^{[d]}_{ij} = \delta_{ij},              & d=0,& \mbox{(no dilution)} \\
{\mathbf P}^{[d]}_{ij} = \delta_{ij}\ \delta_{di},  & d=0,\dots,N-1 & \mbox{(full dilution)}\\
{\mathbf P}^{[d]}_{ij} = \delta_{ij}\ \delta_{d,\, \lfloor Ki/N\rfloor}& d=0,\dots,K-1, & \mbox{(block-$K$)}\\
{\mathbf P}^{[d]}_{ij} = \delta_{ij}\ \delta_{d,\, i\bmod K} & d=0,\dots,K-1, & \mbox{(interlace-$K$)}
\end{array}\]
where $N$ is the dimension of the space of dilution type and $N/K$ is an integer. 
  
The solution vector, $\varphi^{(A)[d]}_{a\alpha jp}$, is then computed by inverting $\Omega=\gamma_4M$ with this source. The solution vector is then smeared with the smearing operator and displaced appropriately,
$$\varphi^{(A)[d]}_{a\alpha jp}({\mathbf x},t)=\left(D_j^{(p)}{\mathbf V}_S{\mathbf V}^\dagger_S{ \Omega}^{-1}\right)_{a\alpha;b\beta}\varrho^{(A)[d]}_{b\beta k\bar{p}}$$ 
where $\Omega=\gamma_4M$ and $M$ is the Dirac operator. The quark propagator (for line $A$) is then given by 
$$\sum_d^{N_d} \varphi^{(A)[d]}_{a\alpha j}({\mathbf x},t)\varrho^{(A)[d]\dagger}_{b\beta k}({\mathbf x_0},t_0).$$
\subsection{Meson Correlation Functions}
Meson correlation functions are formed by correlating meson operators given by
$${\mathcal M}^{[d_1d_2]}_l(\varrho_1,\varphi_2;{\mathbf p},t)=c_{\alpha\beta}^{(l)}\sum_{\mathbf x}e^{-i{\mathbf p}\cdot({\mathbf x}+({\mathbf d}_\alpha+{\mathbf d}_\beta))}\varrho^{[d_1]}_{a\alpha}({\mathbf x},t;\varrho_1)^*\varphi^{[d_2]}_{a\beta}({\mathbf x},t;\varrho_2)$$
where ${\mathbf d}_\alpha$ are introduced to maintain G-parity when needed. The general form of the meson correlator takes the form (many of the indices have been suppressed in this expression for brevity),
\begin{eqnarray}
C_{l\bar{l}}({\mathbf p},t_f-t_0)=&&\left<-\delta_{A\bar{A}}\delta_{B\bar{B}}{\mathcal M}^{[d_1d_2]}_l(\bar{\varphi}_1,\varphi_2;{\mathbf p},t_f){\mathcal M}^{[d_1d_2]}_l(\bar{\varrho}_1,\varrho_2;{\mathbf p},t_0)^*\right.\\\nonumber
&&\left.+\delta_{AB}\delta_{\bar{A}\bar{B}}{\mathcal M}^{[d_1d_1]}_{\bar{l}}(\varrho_1,\varphi_1;{\mathbf p},t_f){\mathcal M}^{[d_2d_2]}_{\bar{l}}(\varphi_2,\varrho_2;{\mathbf p},t_0)^*\right>
\end{eqnarray}
where the second term appears for isoscalar mesons. 

\section{Preliminary Results}
In this section, we present a few selected preliminary results for particles in different isospin sectors. 

\subsection{$I=0$}
\begin{figure}[t]
\begin{center}
\begin{tabular}{cc}
\begin{minipage}{73mm}
\begin{center}
\ig[height=0.25\tht,angle=0]{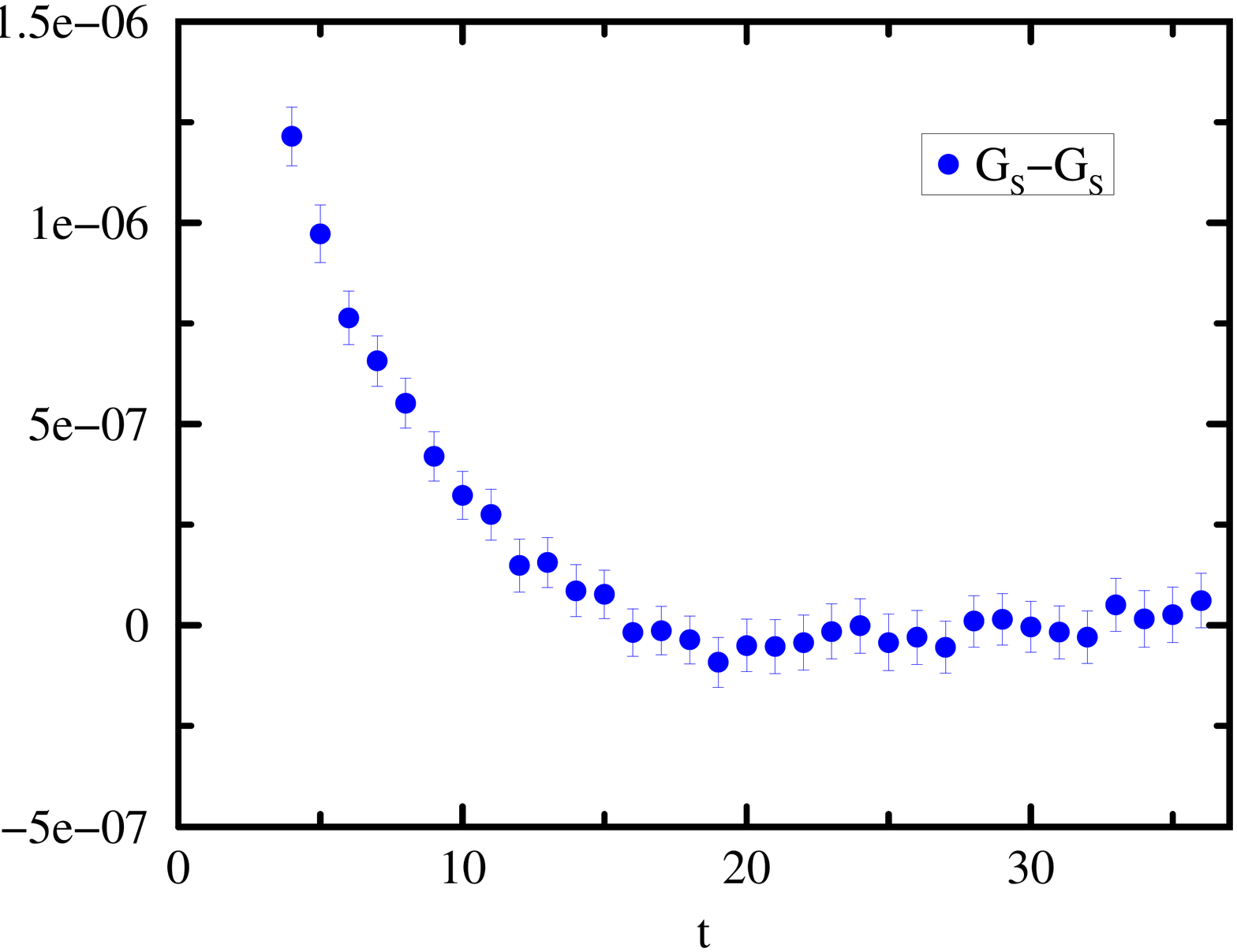}
\caption{Preliminary results for the glueball correlation function on the $24^3$ lattice at $240$ MeV pion mass.}\label{fig:glueball}
\end{center}
\end{minipage}&
\begin{minipage}{73mm}
\begin{center}
\ig[height=0.25\tht,angle=0]{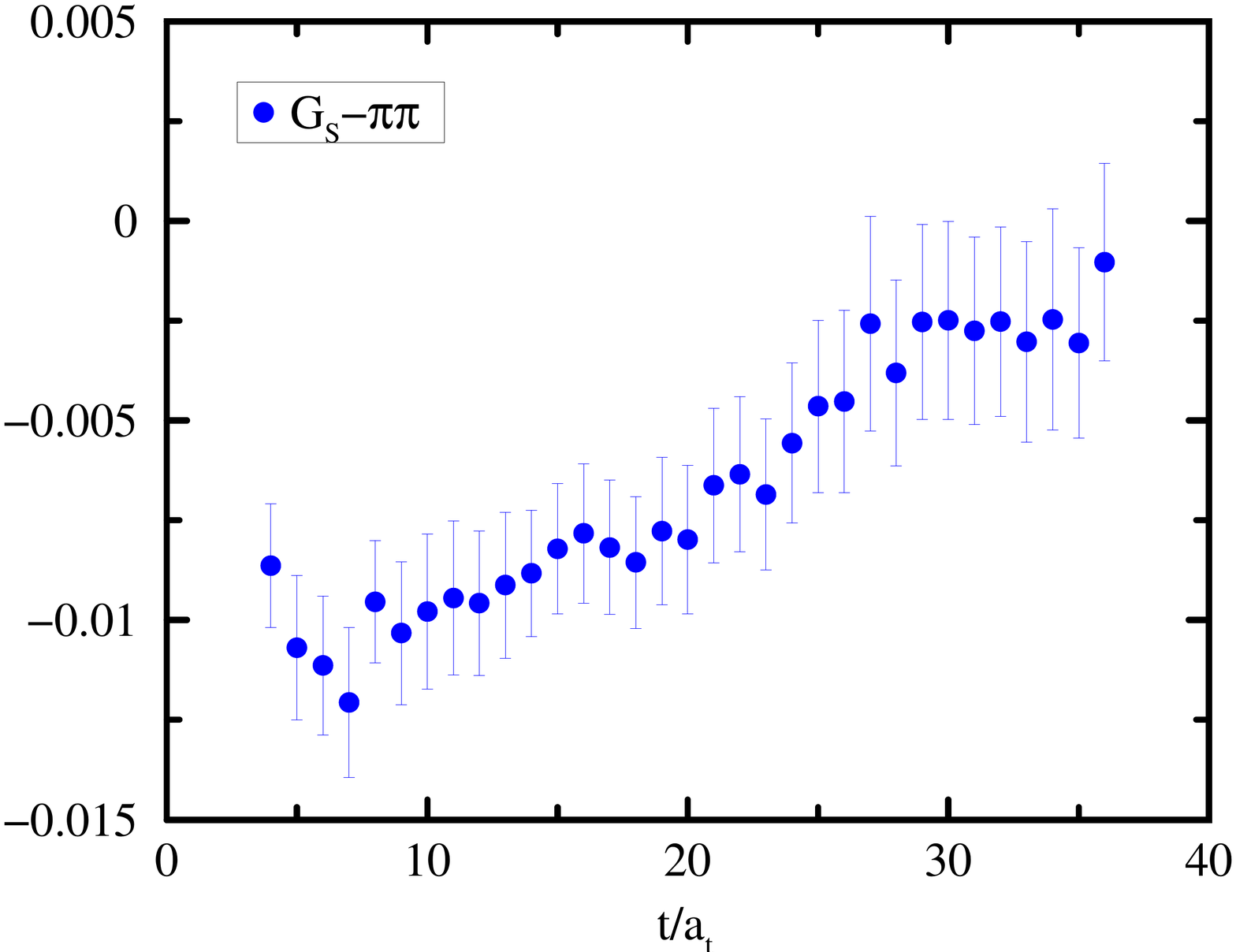}
\caption{Preliminary results for the correlation function of the glueball and $\pi$-$\pi$ mixing term for the 240 MeV configurations.}\label{fig:glueballscalar}
\end{center}
\end{minipage}
\end{tabular}
\end{center}
\end{figure}

\begin{figure}[t]
\begin{center}
\ig[width=0.32\tht,angle=0]{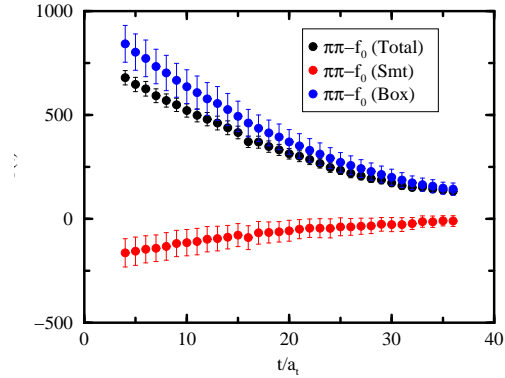}
\caption{Preliminary results for the various contributions to the $\pi\pi-{\rm f}_0$ mixing term.}\label{fig:pipiscalar}
\end{center}
\end{figure}

The isoscalar channel is one of the most challenging channels as there are disconnected (``same-time'') diagrams and ``box'' diagrams appearing in the correlation function. The completely disconnected diagrams generally suffer from large gauge noise and hence is an ideal case for stochastic estimation over `non-stochastic' evaluations such as the distillation algorithm. The ``box diagrams'' are just as difficult to evaluate but are perhaps not as noisy. The stochastic LapH algorithm allows the computation of all of these diagrams which are involved in this channel from four independent noise sources. Two noise sources are needed just to construct a single pion operator and so at the expense of computing two more noise vectors, we can evaluate all of the diagrams needed for the isoscalar channel. The scalar glueball operator comes for free as well because one can construct it from the eigenvalues of the LapH operator. A glueball correlation function using these LapH operators is shown in Fig.~\ref{fig:glueballscalar} on the $24^3\times256$ lattice with the 240~MeV pion mass. 
We have studied the mixings between the scalar meson and the scalar glueball as well as scalar mesons with the $I=0$ $\pi\pi$ scattering state. The off-diagonal correlation functions for these different interpolating operators for the $I=0$ channel on the $24^3$ lattices are shown in Fig.~\ref{fig:glueballscalar} and Fig.~\ref{fig:pipiscalar}. 

\subsection{$I=1/2$}
\begin{figure}[ht]
\begin{center}
\ig[height=0.4\tht,angle=0]{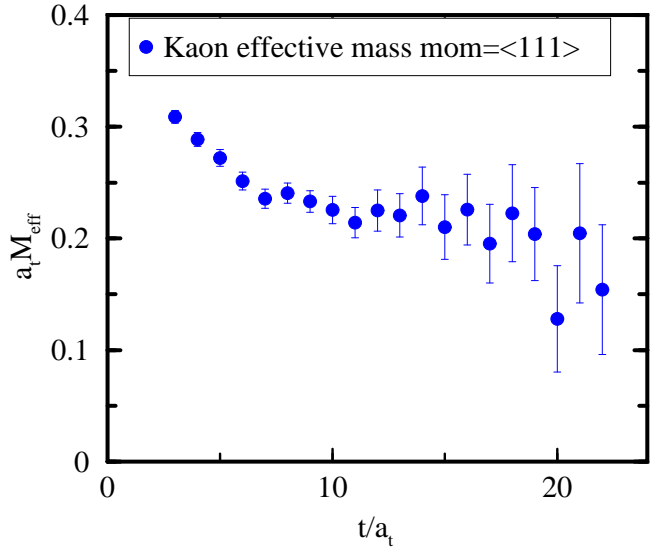} 
\caption{The effective mass of the kaon with momentum $(1,1,1)$ in lattice units.}\label{fig:kaon}
\end{center}
\end{figure}

There are no disconnected diagrams nor any box diagrams that needs to be evaluated for kaons. The gain from the stochastic LapH algorithm is that different momenta values for the kaons can be simulated by introducing simple exponential factors into the zero momentum operators. However, there will be an extra noise source/solution pair needed at the strange quark mass. Our preliminary results suggest that the introduction of the stochastic noise solely in the LapH subspace does not make the determination of energies with finite momentum any harder than zero momentum states with similar energies. We show an example for the kaon with momentum $(1,1,1)$ in lattice units in Fig.~\ref{fig:kaon}.
\subsection{$I=1$}
\begin{figure}[ht]
\begin{center}
\ig[height=0.4\tht,angle=0]{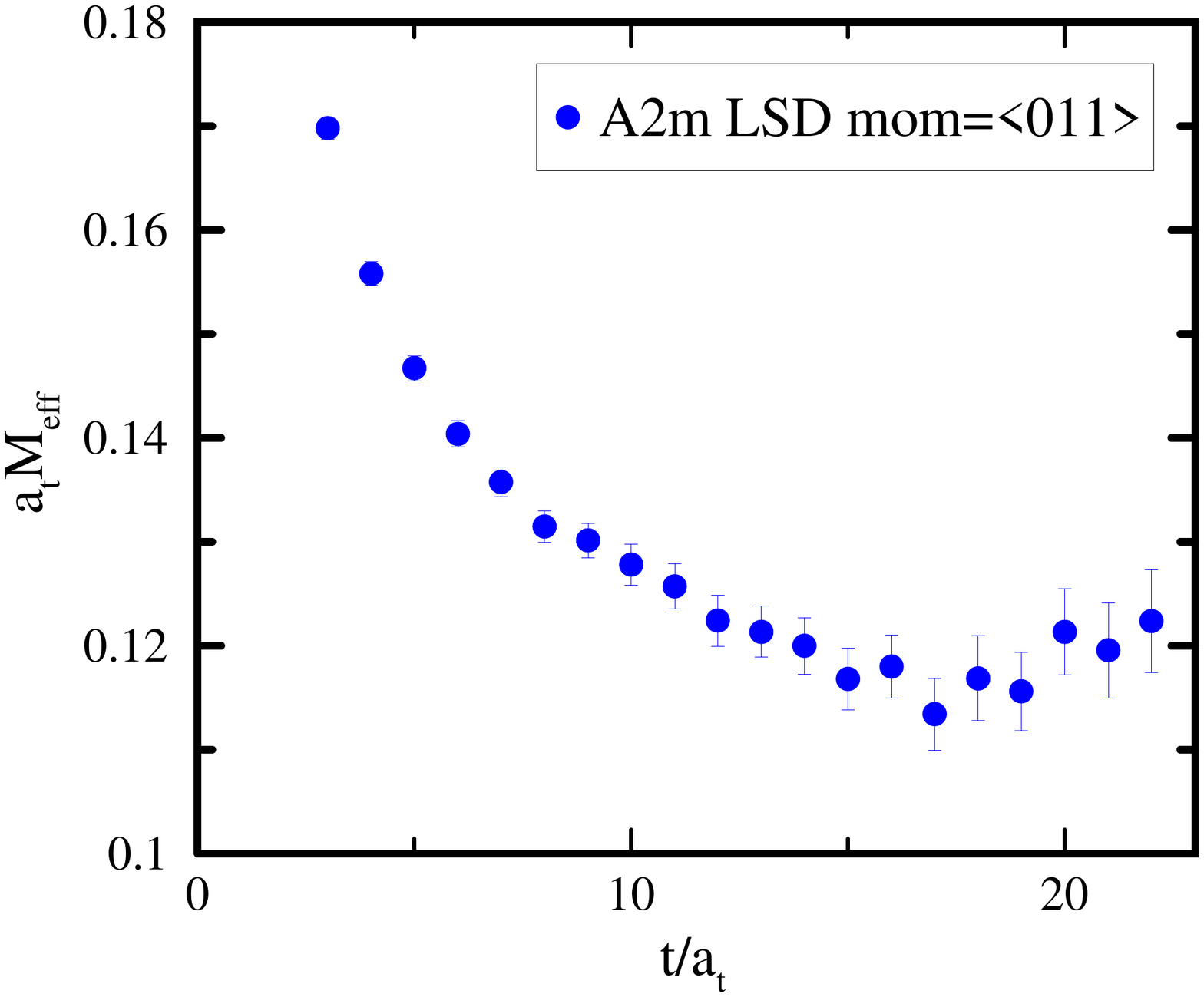} 
\caption{The effective mass of the $\rho$ with momentum $(0,1,1)$.}\label{fig:rho}
\end{center}
\end{figure}

The rho decay width has attracted attention in recent years as several groups have performed large scale simulations at light quark masses (\cite{Aoki:2011yj}-\cite{Feng:2010es}). The lattice sizes and the quark masses we have are limited as larger lattices with lighter masses are still being generated now. We have generated many of the operators with finite momenta that will be needed for the analysis of this channel. We show one of the representative effective masses in this channel in Fig.~\ref{fig:rho}.  

\subsection{$I=2$}
\begin{figure}[t]
\begin{center}
\ig[height=0.4\tht,angle=0]{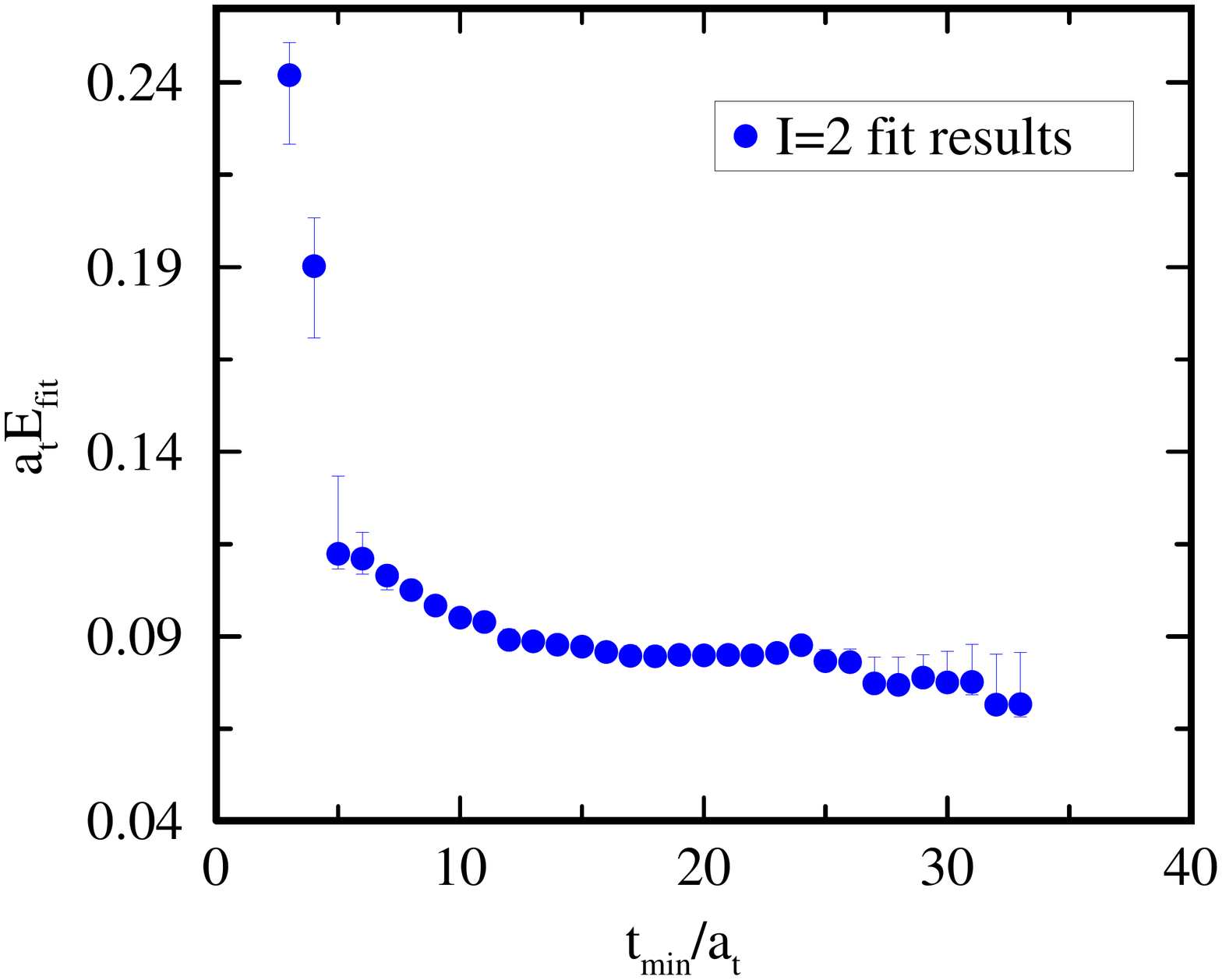} 
\caption{A $t_{min}$ plot of the energies extracted in the $I=2$ channel. }\label{fig:I2}
\end{center}
\end{figure}

The isospin-2 channel is an exotic channel and does not require any disconnected diagrams to be evaluated. The correlation function can be evaluated by computing the product of two pion propagators and the corresponding quark-exchanged diagrams. Since the finite momentum single pion operators have already been computed, we can use many different $\pi\pi$ operators to reduce the contamination from excited states. The use of many operators for this state is important in a large volume because the level spacings decrease with increasing volume. We construct a correlation matrix which is then diagonalized to find the optimized correlation function. We fit the optimal correlator for various $t_{min}$ values to check for stability in the fits. A ``$t_{min}$ plot'' is shown in Fig.~\ref{fig:I2} after diagonalization. The scattering length that is extracted from the fits are consistent with continuum two-loop chiral perturbation theory predictions \cite{Colangelo:2001df}, but the value is sensitive to the choice of the optimization timeslice. We are examining the systematic errors of choosing different diagonalization times and also the sensitivity to $t_{max}$ included in the fits. 

\section{Summary}
The stochastic LapH quark smearing algorithm has been tested for single particle meson states with various momenta and isoscalar channels including the mixing of the $\pi$-$\pi$ state with the scalar glueball and the scalar meson state. The elastic scattering length was estimated using L\"uscher's formula and found to be consistent with large errors with continuum chiral perturbation theory. The $I=2$ exotic state was computed using various finite momenta pion operators and a result consistent with chiral perturbation theory was obtained. The systematic errors due to the different diagonalization procedure has not been fully understood yet and is a subject of current investigation.

\section*{Acknowledgements}
This work has been partially supported by National Science Foundation awards PHY-0510020, PHY-0653315, PHY- 0704171, PHY-0969863, and PHY-0970137, and through TeraGrid/XSEDE resources provided by the Pittsburgh Supercomputer Center, the Texas Advanced Computing Center, and the National Institute for Computational Sciences under grant numbers TG-PHY100027 and TG-MCA075017.

\end{document}